\documentclass[9pt,twocolumn,twoside]{osajnl}

\journal{optica} 

\setboolean{shortarticle}{false}
\usepackage{color}

\title{Single-photon imaging over 200 km}

\author[1,2,3,$\dagger$]{Zheng-Ping Li}
\author[1,2,3,$\dagger$]{Jun-Tian Ye}
\author[1,2,3,$\dagger$]{Xin Huang}
\author[1,2,3]{Peng-Yu Jiang}
\author[1,2,3]{Yuan Cao}
\author[1,2,3]{Yu Hong}
\author[1,2,3]{Chao Yu}
\author[1,2,3]{Jun Zhang}
\author[1,2,3]{Qiang Zhang}
\author[1,2,3]{Cheng-Zhi Peng}
\author[1,2,3,*]{Feihu Xu}
\author[1,2,3,*]{Jian-Wei Pan}

\affil[1]{Hefei National Laboratory for Physical Sciences at Microscale and Department of Modern Physics, University of Science and Technology of China, Hefei 230026, China.}
\affil[2]{Shanghai Branch, CAS Center for Excellence in Quantum Information and Quantum Physics, University of Science and Technology of China, Shanghai 201315, China}
\affil[3]{Shanghai Research Center for Quantum Sciences, Shanghai 201315, China}
\affil[$\dagger$]{These authors contributed equally to this work.}
\affil[*]{Corresponding author: feihu.xu@ustc.edu.cn; pan@ustc.edu.cn}

\dates{Compiled \today}

\ociscodes{(110.1758) Computational imaging; (010.3640) Lidar; (100.6890) Three-dimensional image processing; (270.0270) Quantum optics; (040.3780) Low light level.}

\begin{abstract}
Long-range active imaging has widespread applications in remote sensing and target recognition. Single-photon light detection and ranging (LiDAR) has been shown to have high sensitivity and temporal resolution. On the application front, however, the operating range of practical single-photon LiDAR systems is limited to about tens of kilometers over the Earth's atmosphere, mainly due to the weak echo signal mixed with high background noise. Here, we present a compact coaxial single-photon LiDAR system that is capable of realizing 3D imaging at up to 201.5 km. This is achieved by using high-efficiency optical devices for collection and detection, and a new noise-suppression technique that is efficient for long-range applications. We show that photon-efficient computational algorithms enable accurate 3D imaging over hundreds of kilometers with as few as 0.44 signal photon per pixel (PPP). The results represent a significant step towards practical low-power LiDAR over extra-long ranges.
\end{abstract}

\setboolean{displaycopyright}{true}

\begin{document}

\maketitle

\begin{figure*}[!t]\center
\resizebox{15.5cm}{!}{\includegraphics{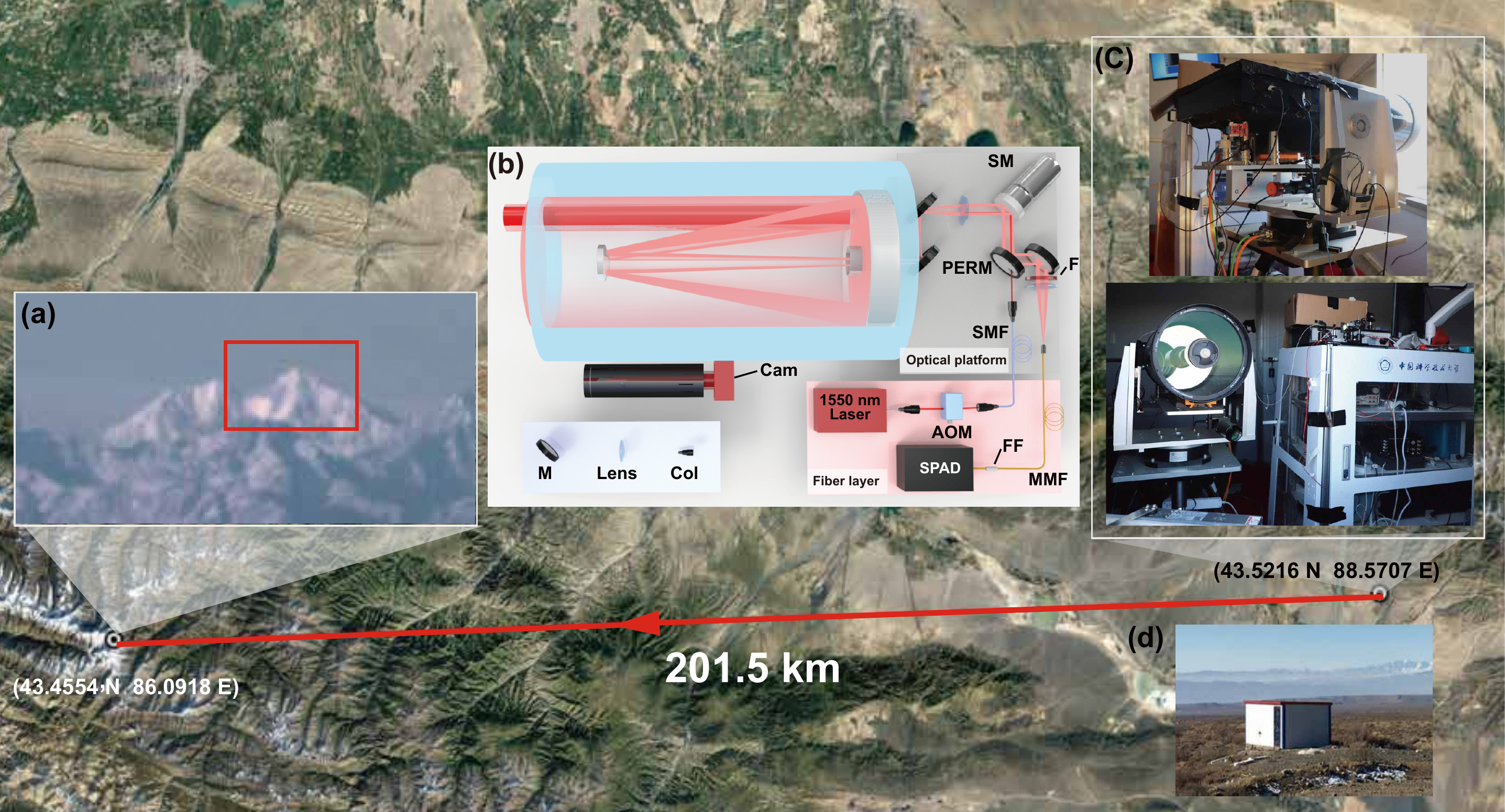}}
\caption{\textbf{Illustration of the long-range active imaging over 201.5 km.} Satellite image of the experiment implemented near the city of Urumqi, where the single-photon LiDAR is placed at a temporary laboratory in the wild. \textbf{(a),} the visible-band photograph of the mountain taken by a standard astronomical camera equipped with a telescope. The elevation of the mountains is approximately 4500 m.  \textbf{(b),} Schematic diagram of the experimental setup. SM, scanning mirror; Cam, camera; M, mirror; PERM, $45^o$ perforated mirror; SPAD, single-photon avalanche diode; MMF, multimode fiber; Col, collimator; AOM, acoustic-optic modulator; F, spectral filter; FF, fiber spectral filter; SMF, single mode fiber. \textbf{(c),} Photograph of the setup hardware, including the optical system (top and bottom left) and the electronic control system (bottom right). \textbf{(d),} the view of the temporary laboratory at an altitude of 1770 m. The elevation profile of the terrain between the setup and the targets is shown in Supplementary Fig. S1.}
\label{Fig:Fig1}
\end{figure*}

\section{Introduction}
Single-photon LiDAR~\cite{degnan2001unified,degnan2002photon,marino2005jigsaw} based on TCSPC~\cite{hadfield2009single} can provide single-photon sensitivity and picosecond resolution for time-of-flight measurements~\cite{buller2007ranging}, and it has foreseen significant progress in the experimental developments for various applications~\cite{zappa2007principles,sun20133d,villa2014cmos,gariepy2015single,laurenzis2007long,mccarthy2009long,mccarthy2013kilometer,zhou2015few,laurenzis2015multiple,li2017multi,pawlikowska2017single,li2020single,ren2018high,chan2019long,tachella2019real,li2020super} (see~\cite{altmann2018quantum} for a review). Single-photon LiDAR was originally proposed for laser rangers and altimeters~\cite{degnan2001unified,degnan2002photon}, which have been transformative for long-range laser ranging~\cite{dickey1994lunar} and global topography~\cite{neuenschwander2019atl08}. Geiger-mode single-photon LiDAR systems, originally developed by MIT Lincoln Laboratories~\cite{marino2005jigsaw}, have been widely adopted for airborne and spaceborne topographic measurements~\cite{glennie2013geodetic}. These techniques have been made commercially available (e.g., Sigma Space/Hexagon) for fast measurements with single-photon detector arrays. Single-photon LiDAR built in high-altitude airborne platforms can provide continuous topographic and bathymetric mapping~\cite{degnan2016scanning}. For large-scale topography, satellite-based laser altimeter, such as ATLAS on ICESat-2~\cite{neuenschwander2019atl08}, has been adopted for long-term observations of earth surface.

In recent years, single-photon LIDAR for long-range imaging over the terrestrial (Earth's) atmosphere has received a lot of research attention~\cite{laurenzis2007long,mccarthy2009long,mccarthy2013kilometer,villa2014cmos,zhou2015few,laurenzis2015multiple,li2017multi,pawlikowska2017single,li2020single,ren2018high,chan2019long,tachella2019real,li2020super}, since it can provide high-resolution three dimensional (3D) imaging in both transversal and longitudinal dimension (at centimeter scale). This capability plays important roles for target recognition and identification over long ranges. Meanwhile, advanced computational algorithms have permitted 3D imaging with a small number of photons, i.e., one photon per pixel~\cite{kirmani2013first,altmann2016lidar,shin2016photon,shin2015photon,rapp2017few,lindell2018single,tachella2019bayesian,peng2020photon}. The combination of the single-photon LiDAR and photon-efficient imaging algorithms has witnessed the realization of long-range imaging at tens of kilometers~\cite{pawlikowska2017single,li2020single}. Despite the remarkable progress, further extending the distance remains challenging. In long-range scenarios along the Earth's atmosphere, the number of echo signal photons from the scene of interest decreases quadratically with the distance~\cite{wagner2006gaussian}, while the background noise (mainly attributing to the backscatters from near-field atmosphere) is linearly correlated with the output laser power. Even with a high-power laser, the limited signal-to-background ratio (SBR) in long ranges prevents useful reconstructions. Based on state-of-the-art algorithms~\cite{rapp2017few,lindell2018single}, the previous single-photon imaging experiments over the terrestrial atmosphere~\cite{pawlikowska2017single,li2020single} are typically limited to an achievable range below one hundred kilometers (see Supplementary Table).

Here we demonstrate a high-efficiency and low-noise coaxial single-photon LiDAR system that is capable of realizing 3D imaging at up to 201.5 km with as few as 0.44 signal photon per pixel (PPP). This is achieved by the developments of an optimized transceiver optics, low-noise InGaAs/InP single-photon avalanche diode detector (SPAD) and an efficient noise-suppression technique. The validity of different photon-efficient imaging algorithms~\cite{shin2015photon,rapp2017few,lindell2018single,tachella2019bayesian} has been verified under this extremely low-light-flux condition. Furthermore, we demonstrate photon-efficient laser ranging over one hundred kilometers, and show that the range of non-cooperative targets can be accurately measured with 4.02 signal photon counts only. The demonstrations of imaging and ranging with a small number of photons over 200-km range are desirable for the general applications of target recognition and satellite-based topography.

\section{Single-photon LiDAR setup}
As shown in Fig.~\ref{Fig:Fig1}, the single-photon LiDAR system primarily utilizes commercial off-the-shelf devices that operate at room temperature, where the source is a standard fiber laser and the detector is a compact low-noise InGaAs/InP SPAD. The transceiver system consists of a commercial Cassegrain telescope with a modest telescope aperture of 280 mm and a custom-built integrated optical platform mounted on a homemade two-axis rotating stage. To obtain high atmospheric transmittance and low solar background in eye-safe wavelengths, we choose a near-infrared (NIR) wavelength of 1550 nm. For the source, we employ an all-fiber pulsed erbium doped fiber laser (600 ps pulse width and 500 kHz clock rate),where micro joule pulses (with a maximum laser power of 600 mW) and an expanded beam design (with a diameter of 7 cm for the beam emitted from the telescope) are adopted to meet the eye-safety standards~\cite{standard2000american}, the systems falls within laser eye-safety thresholds. The laser is coupled to a single-mode fiber before transmission in order to encompass the divergence for illumination.

To balance the ambient light rejection and coupling efficiency, the returned photons are coupled into a multimode fiber with an appropriate core diameter (62.5 $\mu m$) for high-efficiency collection. Similar to previous single-photon LiDAR systems~\cite{mccarthy2009long,mccarthy2013kilometer,pawlikowska2017single}, our setup uses a coaxial scanning design (see Fig.~\ref{Fig:Fig1}) for transmitting and receiving optical paths, rather than conventional dual-telescope configuration~\cite{neuenschwander2019atl08}. This configuration has the capability of precisely aligning the transmitting and receiving spots projected on the target over dynamical distances, and facilitating fine scanning to achieve high-resolution imaging. High-precision scanning is implemented by closed-loop coplanar dual-axis piezo tip-tilt platform in both $x$ and $y$ axial directions.

\begin{figure*}[!t]\center
\resizebox{16.3cm}{!}{\includegraphics{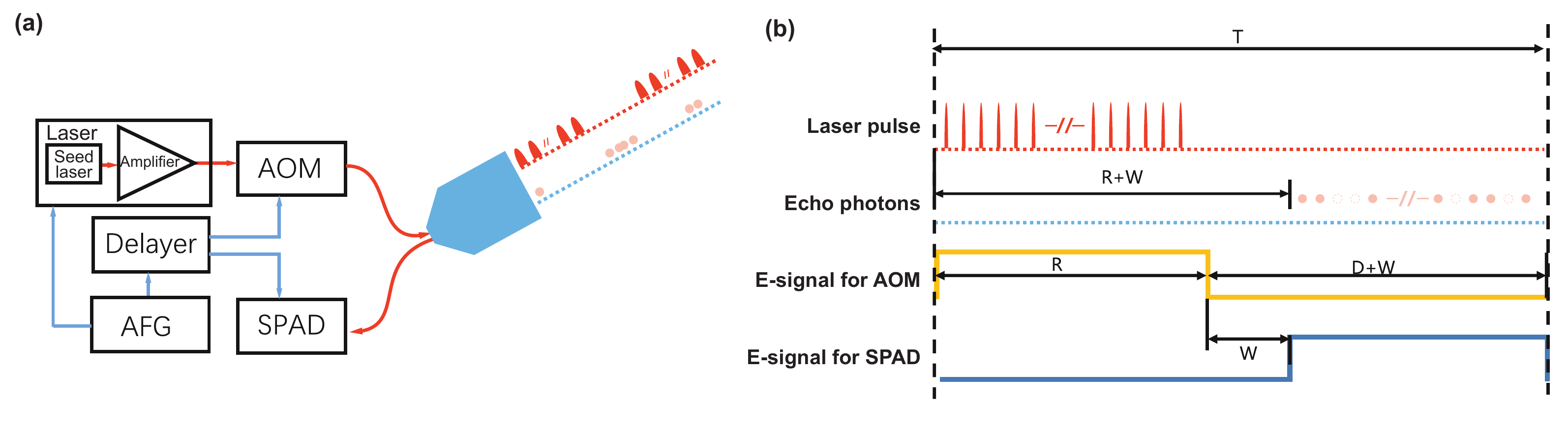}}
\caption{\textbf{Schematic diagram of the noise-suppression technique.} \textbf{(a),} The arbitrary function generator (AFG) triggers the laser and the delayer. An acoustic-optic modulator (AOM) is used to block the ASE noise when the laser is triggered off. A Delayer provides electronic signal (E-signal) to the SPAD and AOM. \textbf{(b),} The timing diagram in an operational period $T$. During the emission mode $R$, the laser pulses are emitted, and the AOM is on to allow pulse emission. At the end of $R$, the AOM is switched off to block ASE noise. After an isolation period $W$, the SPAD is gated on to detect the back-reflecting singal photons for a detection period of $D$.}
\label{Fig:Fig2}
\end{figure*}

\begin{figure}[h]\center
\resizebox{8.5cm}{!}{\includegraphics{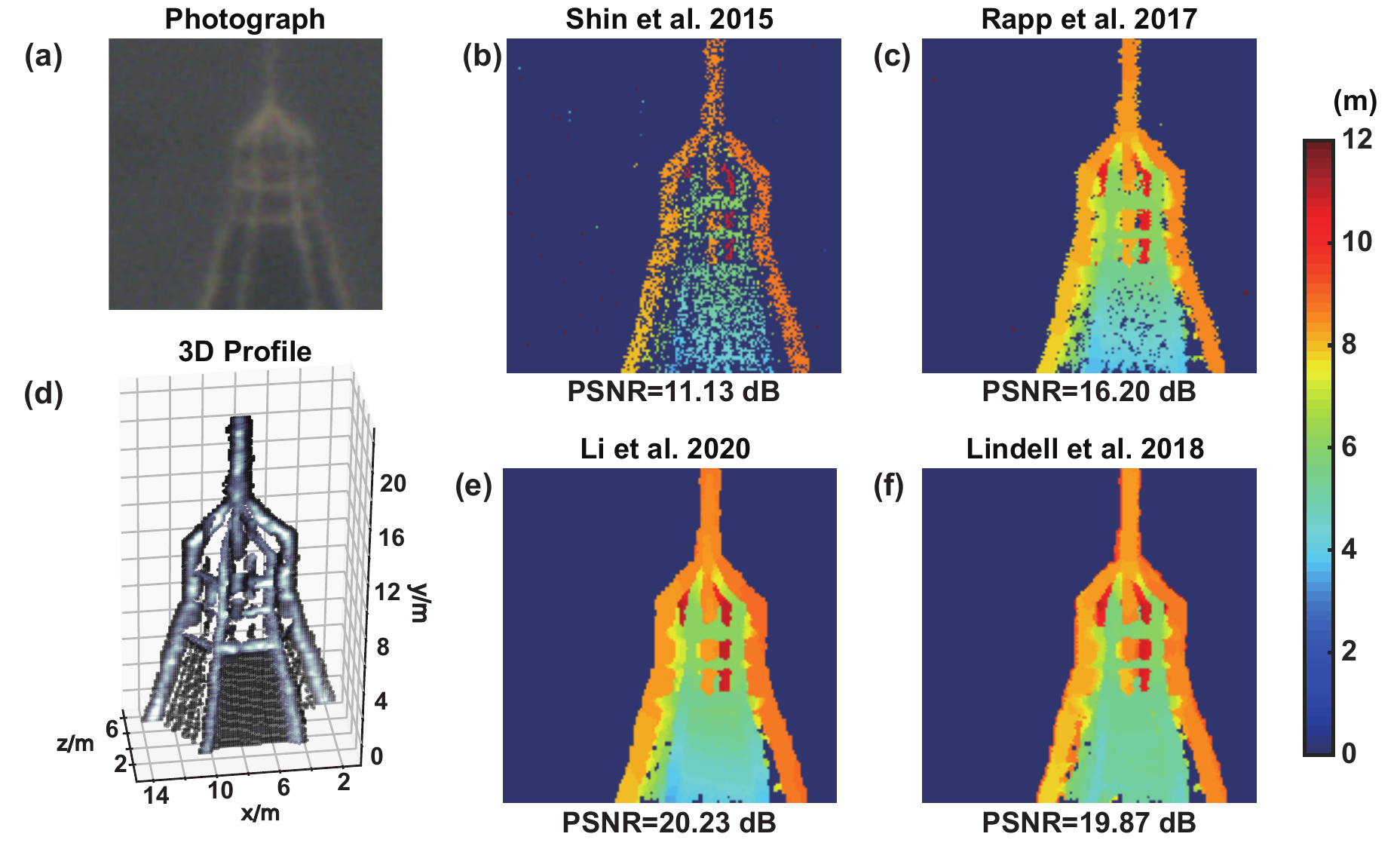}}
\caption{\textbf{Reconstruction results for a tower over 9.8 km.} \textbf{(a),} real visible-band photo taken with a standard astronomical camera. \textbf{(b),(c),(e),(f),} the reconstructed depth results by different photon-efficient algorithms, including Shin et al.\cite{shin2015photon}, Rapp et al.\cite{rapp2017few}, Li et al.\cite{li2020single}, and Lindel et al.\cite{lindell2018single}. \textbf{(d),} 3D demonstration of the reconstructed result from Lindell's method. The SBR is $\sim$15.76 and the mean signal PPP is $\sim$3.47. Note that a groundtruth is generated by Lindell's method from a dataset with 35 signal PPP for the PSNR's calculation.}
\label{Fig:Fig8}
\end{figure}

\section{Optical techniques tailored for long ranges}

To facilitate 3D imaging over long ranges beyond tens of kilometers, we develop high-efficiency optical devices and an efficient noise-suppression method in order to improve the collection efficiency and decrease the background noise. The SBR of our system is much better than the previous works~\cite{pawlikowska2017single,li2020single} (a detailed comparison of the performance is shown in Supplementary Table.~S1 and Fig.~S4).

\subsection{Improving the collection efficiency.}
First, we develop an InGaAs/InP SPAD~\cite{Yu2017Fully,korzh2014free} with detection efficiency of 19.3$\%$, time jitter of$\sim$180 ps, and dark count rate as low as 0.1 KHz. We use a lightweight (2 kg) and low power (peak power 60 W) thermoacoustic cooler to cool negative feedback avalanche diode (NFAD) devices down to 173 K in order to achieve low dark count rate (DCR)~\cite{Yu2017Fully}. Moreover, the NFAD is coupled by a multi-mode fiber with a core diameter of 62.5 $\mu m$ to enhance collection efficiency. In experiment, the SPAD is optimized to $\sim$100 cps DCR at 19.3\% PDE. Because of the low temperature for the SPAD, the DCR is about 20 times lower than previous long-range single-photon LiDAR systems~\cite{pawlikowska2017single,li2020single}. Second, the telescope is coated to achieve high transmission at 1550 nm. The coating materials including $SiO_2$ and $Ga_2O_5$ are accumulated on lenses by means of ion-beam assisted deposition. As compared to the ordinary commercial telescope (which is coated for visible light), the total transmission is increased by two times, i.e., from about 50\% to 95\%. Finally, the transmitting divergence angle and the receiving field of view (FoV) are designed to be 17.8 $\mu rad$ and 11.2 $\mu rad$ respectively. These angles are twice smaller than previous work~\cite{pawlikowska2017single,li2020single}, and they are close to the diffraction limit of the telescope aperture. Note that the smaller FoV can reduce the ambient light by a factor of 75\%, but provides increased higher resolution. Moreover, it allows narrow photon clusters in temporal domain, thus mitigating the issue of multiple returns~\cite{hernandez2007bayesian,tachella2019bayesian,li2020single}.
\begin{figure}[!b]\center
\resizebox{8.5cm}{!}{\includegraphics{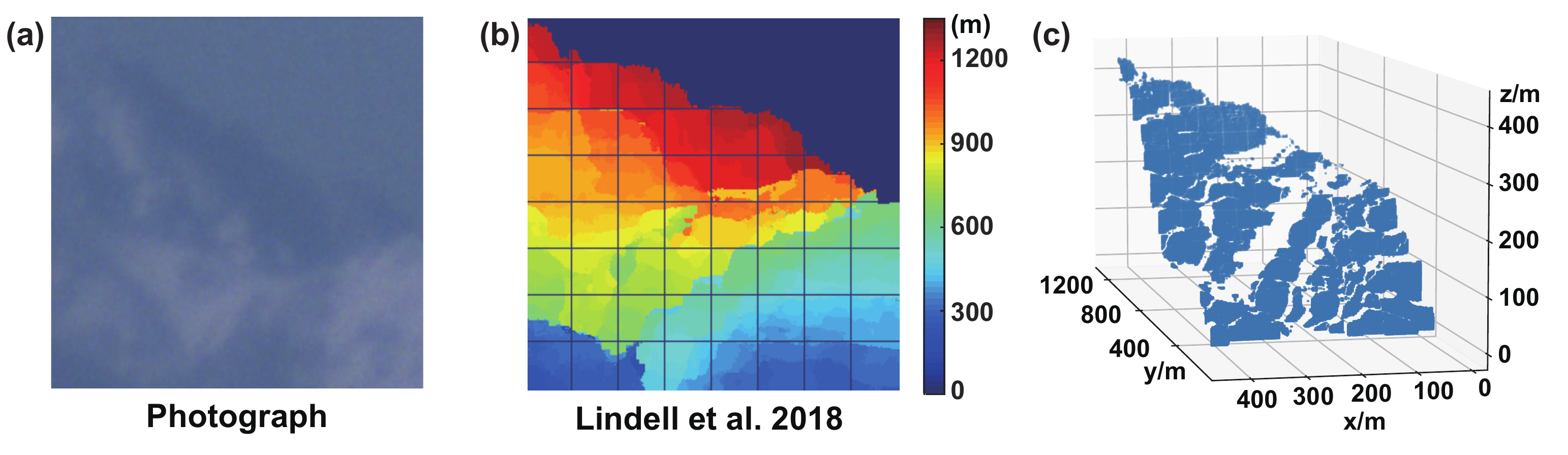}}
\caption{\textbf{Reconstruction results for a scene over 124.0 km.} \textbf{(a),} real visible-band photo. \textbf{(b),} the reconstructed depth result by Lindell et al 2018 \cite{lindell2018single} for the data with SBR$\sim$0.51 and mean signal PPP$\sim$3.86. \textbf{(c),} a 3D profile of the reconstructed depth map.}
\label{Fig:Fig3}
\end{figure}

\subsection{Noise-suppression technique.}
In long-range LiDAR, the number of echo signal photons is highly limited~\cite{kirmani2013first,altmann2016lidar,shin2016photon,shin2015photon,rapp2017few,lindell2018single}, which imposes a stringent requirement for the system's background noise level. By using the low-noise SPAD, we characterize the background noise and find that the noise mainly comes from two types of backscattering noise. The first one is the backscatter of the laser pulse from the near-field atmosphere and the common transmitting/receiving optics elements. Particularly, the atmospheric scattering lasts a long period after each pulse emission. The second one is the backscatter of the amplified spontaneous emission (ASE) noise. The ASE noise, which is unavoidable due to the required optical amplification, occurs in the full temporal domain of the detection (see Supplementary for details). These two types of noise can \textbf{not} be removed by a conventional gating operation of the SPAD~\cite{pawlikowska2017single,mccarthy2009long,mccarthy2013kilometer}.

\begin{figure*}[btp]\center
\resizebox{15.7cm}{!}{\includegraphics{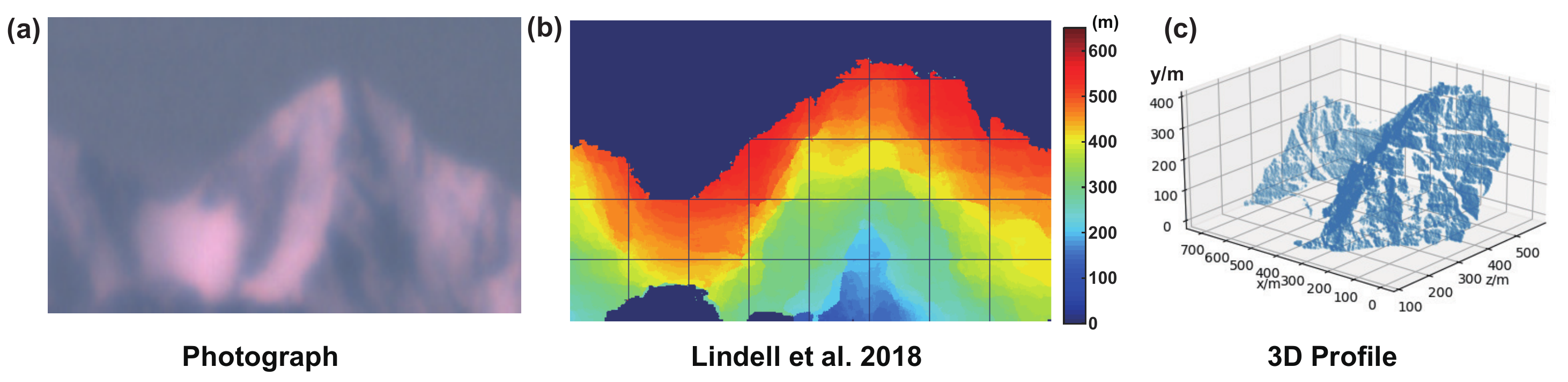}}
\caption{\textbf{Reconstruction results of a scene over 201.5 km.} \textbf{(a)}, real visible-band photo. \textbf{(b),} the reconstructed depth result by Lindell et al 2018 \cite{lindell2018single} for the data with SBR$\sim$0.04 and mean signal PPP$\sim$3.58. \textbf{(c),} a 3D profile of the reconstructed result.}
\label{Fig:Fig4}
\end{figure*}

In contrast, we developed an efficient temporal filtering approach for noise suppression. In our approach shown in Fig.~\ref{Fig:Fig2}, we set a temporal separation of the emission mode ($R$) and the detection mode ($D$), and employ an additional high-extinction-ratio acoustic optical modulator (AOM) to realize fast switching between these two modes. In the emission mode, the laser pulses are triggered at a high repetition rate, while the SPAD is turned off using electrical gating; in the detection mode, the laser stops emitting pulses and the SPAD is turned on. This can eliminate the local noise from the pulse emission (originally 2.4 nW coupled into the SPAD), and reduce the noise by a factor of 100. Note that even when the laser pulses are not triggered during the detection mode, the optical amplifier will still contribute ASE noise ($\sim10^9$ photons/s coupled). Therefore, we use the AOM to efficiently isolate the ASE noise with an extinction ratio of about 57 dB. We also set an transition time of $W$ in Fig.~\ref{Fig:Fig2} between emission mode and detection mode to further subside near-field atmospheric reflection. For phase settings (R, W, D), taking 200 km imaging as an example, since the round-trip flight time of photons is 1.34 ms,  we set the emission mode at period [0, 1.2 ms] and the detection mode at [1.3 ms, 2.5 ms] to achieve an optimized efficiency.

The total number of noise photon counts is quantified to be about 0.4 KHz, which is at least 50 times smaller than previous works~\cite{pawlikowska2017single,li2020single} (see Supplementary Fig.~S3). This enhancement is mainly due to three factors: the time gating approach, the low noise SPAD and the smaller FoV. The ultra-low-noise is the key enabling feature to image and range over hundreds of kilometers.

\section{Results}
Using our system, we perform an in-depth study to image a variety of natural scenes over long ranges. Most of the experiments were done in the wild environment at night near the city of Urumqi, China. Note that we also have the ability to image in daytime (see Supplementary Fig. S2). In experiment, we perform blind LiDAR measurements without any prior information of the absolute time location of returned signals, where the SPAD is free-running in detection mode. Our imaging captures the relative depth of scenes, while standard laser ranging measures the absolute distance. Depth maps of the targets up to 201.5 km were reconstructed with $\sim$1 PPP for signal photons and a SBR as low as 0.04. In experiment, the emission laser power is fixed at 600 mW, and the targets are raster-scanned at different dwell times (depending on the imaging ranges) to form the 3D images. In the following, we present three representative results and show further results in Supplementary Fig.~S3.

First, to demonstrate the capability of high-resolution 3D imaging, we image a tower head over a range about 9.8 km with 160$\times$160 pixels and an acquisition time of 1.5 ms per pixel. The results are shown in Fig.~\ref{Fig:Fig8}. Our setup permits a rather high SBR up to 15.76 (31.80) for the whole picture (on those non-empty pixels), which is about 50 times higher than previous results~\cite{pawlikowska2017single,li2020single}. The fine details of the target can be clearly reconstructed with a signal level of about 3.47 (7.01) signal PPP. With the experimental data, we verify various photon-efficient imaging algorithms~\cite{shin2015photon,rapp2017few,lindell2018single,li2020single}, as exhibited in  Fig.~\ref{Fig:Fig8}(b)(c)(e)(f), where the quantitative performances for each algorithm are given with peak-signal-to-noise-ratio (PSNR). The results show that despite different performances, most of the algorithms are effective to resolve the 3D shape even under the low-light level. For the algorithms based on convex optimization~\cite{shin2015photon,rapp2017few,li2020single}, the parameters are the same as the ones reported in their paper. To make a fair comparison, the regularizer (TV norm) is set to 0.1. For the learning approach~\cite{lindell2018single}, we import the pre-trained CNN directly, and apply it to the raw data. The processing times by a standard laptop for these algorithms are 39.60 s~\cite{shin2015photon}, 77.54 s~\cite{rapp2017few}, 112.05 s~\cite{li2020single} and 9.60 s~\cite{lindell2018single} respectively.


Next, we choose a scene of mountains 124.0 km away. Fig.~\ref{Fig:Fig3}(a) shows the visible-band photograph taken by a standard astronomical camera (ASI294MC) equipped with a telescope. We scan 256$\times$256 pixels with an acquisition time of 32.0 ms per pixel, and obtain a SBR of 0.33 (0.51) and an averaged signal PPP of 2.91 (3.86) for the whole picture (or those non-empty pixels). A depth map reconstructed from~\cite{lindell2018single} and a 3D profile are shown in Fig.~\ref{Fig:Fig3}(b)(c).

\begin{figure}[!h]\center
\resizebox{8.5cm}{!}{\includegraphics{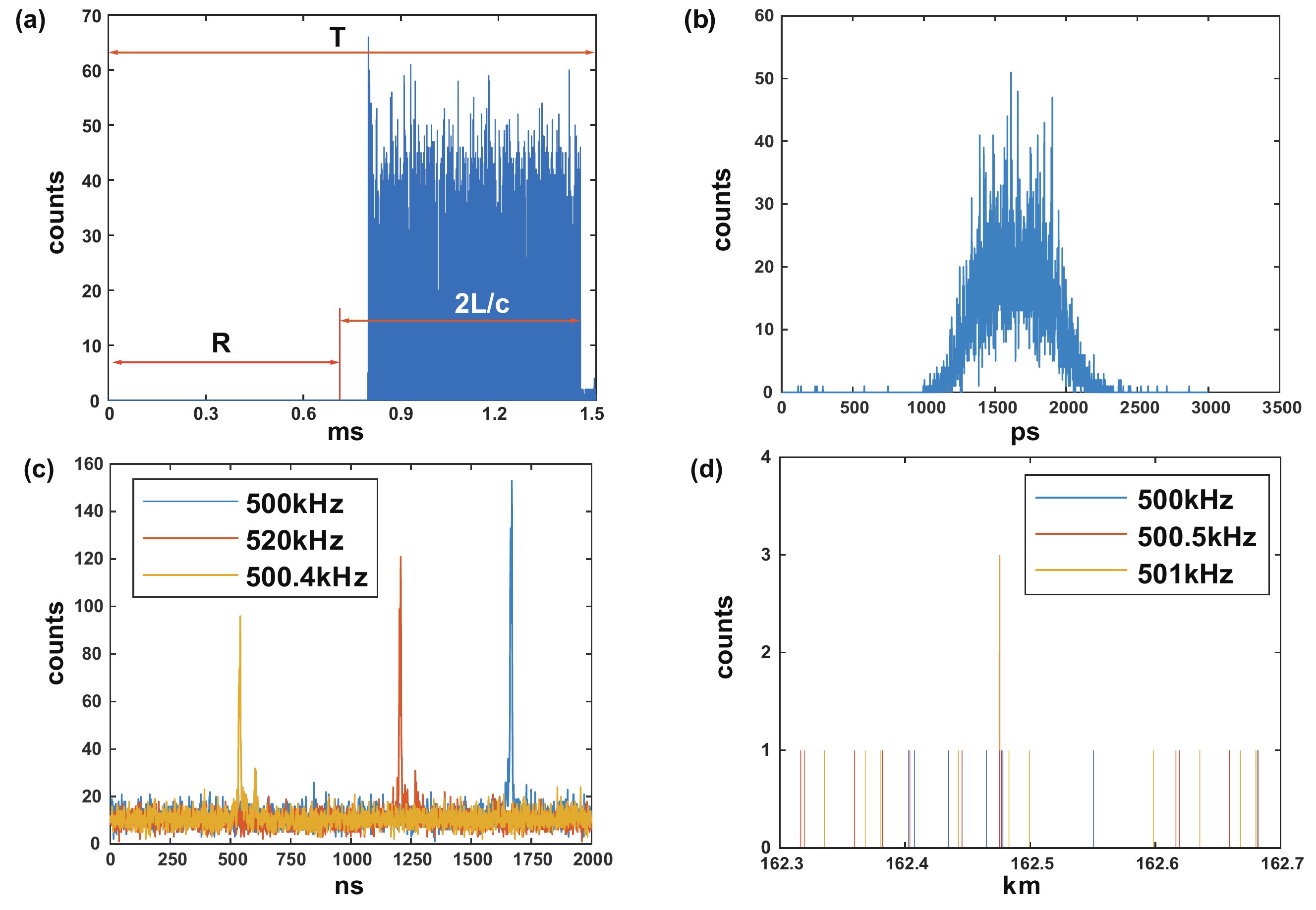}}
\caption{\textbf{Ranging over a hundred kilometers.} \textbf{a,} histograms of raw data from a retro-reflector at absolute distance $L$ = 113.6 km with $T=1.5~ms$ and $R=0.7~ms$. \textbf{b}, the shape of echo signal peak, from which the system jitter can be assessed as FWHM = 600 $ps$. \textbf{c}, histogram of the ranging experiment over 163.3 km with three different repetition rates for a non-cooperation scene of mountains. \textbf{d}, histogram of the measured signals in the photon-efficient ranging experiment, where the averaged number of signal photon counts is about 4.04.}
\label{Fig:Fig7}
\end{figure}

To demonstrate long-range imaging, we choose another target 201.5 km away. We did the imaging in a weather with high visibility of about one hundred kilometers. Fig.~\ref{Fig:Fig4}(a) shows the visible-band photograph. We raster-scan the target with 320$\times$512 pixels at an acquisition time of 189.7 ms per pixel. For the observed data, the SBR is 0.03 (0.04) and the averaged number of signal PPP is 2.23 (3.58) for the whole picture (or those non-empty pixels). A depth map reconstructed by~\cite{lindell2018single} and a 3D profile is shown in Fig.~\ref{Fig:Fig4}(b)(c). In Supplementary Fig.~S2, we further show that our LiDAR system can resolve the 3D details of the long-range scene even with 0.44 signal PPP.

Note that the reconstructed depth imaging is the relative depth information. For complementary purpose, our LiDAR system also has the capability for long-range laser ranging\cite{wilkinson2019next,dickey1994lunar,degnan2008laser,degnan2002asynchronous}, which can provide accurate absolute distances. To show this capability, we demonstrate the laser ranging over hundred kilometers terrestrial atmosphere. LiDAR systems with a high-repetition-rate laser are favored for laser ranging~\cite{buller2007ranging}, but the short laser period will cause the range ambiguity for long-range applications. Previous works have proposed solutions for ranging over tens of kilometers~\cite{du2018high,krichel2010resolving}. Using our low-noise single-photon LiDAR system, we report accurate laser ranging over hundred kilometers with a small number of signal returns. This is achieved by the measurements with multiple repetition rates~\cite{liang20141550} and a photon-efficient approach for distance estimation (see Supplementary). Fig.~\ref{Fig:Fig7}(a) shows the results for laser ranging of a retro-reflector over 113.6 kilometers, where the system's timing jitter is characterized to be 600 $ps$ full width at half maximum (FWHM) as shown in Fig.~\ref{Fig:Fig7}(b). In Fig.~\ref{Fig:Fig7}(c), we show the ranging results with three different repetition rates for a non-cooperation scene of mountains. With an acquisition time of 1.5 s for each measurement and a total number of 30 measurements, we get the absolute distance of 163.337 km with an precision of 3.5 cm. This precision is mainly due to the large FoV which may cover non-smooth multiple depths in each measurement (see Supplementary). Finally, in Fig.~\ref{Fig:Fig7}(d), we demonstrate the photon-efficient ranging by measuring a mountain top with a total number of$\sim$4.04 signal photon counts only (within a total acquisition time of 50 ms). The absolute distance is measured to be 162.476 km. This result shows that our system is capable of laser ranging at 20 Hz over a hundred kilometers using a small number of photons only.

\section{Discussion and Conclusion}

Although the space-borne laser altimetry based on single-photon LiDAR, like the ATLAS on the ICESAt-2~\cite{neuenschwander2019atl08}, has been reported, it is operated with large field of view (FoV) and wide separation between laser beams. However, we focus on single-photon LiDAR for target recognition along terrestrial atmosphere, where one important consideration is high imaging resolution. Hence, our system adopts small FoV and fine scanning for high-resolution imaging. Moreover, for satellite-based laser altimeters, the equivalent vertical thickness of atmosphere is about 5-10 kilometers, i.e., most of the light path is vacuum. In contrast, our experiment is operated over Earth's atmosphere, where the main challenges are the strong atmospheric attenuation and ambient noise from the near-field atmospheric backscattering. To conquer these issues, we enhance the system efficiency and adopt a new time-gating scheme to efficiently suppress the background noise (see Section 3), striving to achieve high SBR (see Supplementary Table S1). Furthermore, we demonstrate accurate 3D imaging with a small number of photons, i.e., one photon per pixel. This sensitivity outperforms previous space-borne LiDARs. Nevertheless, the space-borne LiDARs present array of multiple beamlets, which have the advantages of higher data collection rates and easy operations over much longer distances.

To sum up, we experimentally demonstrate single-photon 3D imaging and ranging up to 201.5 km over terrestrial atmosphere with a small number of photons. We have proposed an effective solution for the noise reduction to allow the single-photon LiDAR with a high SBR. Our system adopts a compact design with low-power laser and commercial off-the-shelf components, and our techniques are photon-efficient. The photon-efficient algorithms and the developed low-noise techniques could facilitate the adaptation of the system for the use in future multibeam single-photon LiDAR systems with Geiger-mode SPAD arrays for rapid remote sensing~\cite{marino2005jigsaw,degnan2016scanning}. The developed techniques are applicable to other wavelength, e.g., 532 nm, which permits water penetration capability. Also, by adopting high-power laser and high-efficiency SPAD array, multibeam single-photon LiDAR may be feasible for fast imaging without scanning. Overall, our results may provide enhanced methods for low-power single-photon LiDAR mounted on low-earth-orbit~\cite{neuenschwander2019atl08} or nano satellites, as a complement to traditional imaging, for high-resolution active imaging and sensing over long ranges.

\section*{FUNDING INFORMATION}
This work was supported by National Key Research and Development (R\&D) Plan of China (2018YFB0504300), National Natural Science Foundation of China (61771443,62031024), Anhui Initiative in Quantum Information Technologies, Shanghai Municipal Science and Technology Major Project (Grant No.2019SHZDZX01), Shanghai Science and Technology Development Funds (18JC1414700).

\section*{Disclosures}
The authors declare no competing financial interests.

\section*{Data and code availability}
The processing code and data can be seen in GitHub: https://github.com/quantum-inspired-lidar/Long-range-single-photon-imaging-over-200-km


\end{document}